\documentclass[pre,twocolumn,superscriptaddress,nofootinbib]{revtex4-2}
\usepackage{graphicx,amssymb,amsfonts,amsmath,chemarr,color,commath,braket,hyperref}

\begin{document}

\title{Distinguishable spreading dynamics in microbial communities}

\author{Meiyi Yao}
\affiliation{Department of Physics and Astronomy, University of Pittsburgh, Pittsburgh, Pennsylvania 15260, USA}

\author{Joshua M. Jones}
\affiliation{Department of Physics, Boston University, Boston, Massachusetts 02215, USA}
\affiliation{Department of Biology, Boston University, Boston, Massachusetts 02215, USA}
\affiliation{Biological Design Center, Boston University, Boston, MA 02215, USA}

\author{Joseph W. Larkin}
\affiliation{Department of Physics, Boston University, Boston, Massachusetts 02215, USA}
\affiliation{Department of Biology, Boston University, Boston, Massachusetts 02215, USA}
\affiliation{Biological Design Center, Boston University, Boston, MA 02215, USA}

\author{Andrew Mugler}
\email{andrew.mugler@pitt.edu}
\affiliation{Department of Physics and Astronomy, University of Pittsburgh, Pittsburgh, Pennsylvania 15260, USA}

\begin{abstract}
A packed community of exponentially proliferating microbes will spread in size exponentially. However, due to nutrient depletion, mechanical constraints, or other limitations, exponential proliferation is not indefinite, and the spreading slows. Here, we theoretically explore a fundamental question: is it possible to infer the dominant limitation type from the spreading dynamics? Using a continuum active fluid model, we consider three limitations to cell proliferation: intrinsic growth arrest (e.g., due to sporulation), pressure from other cells, and nutrient access. We find that memoryless growth arrest still results in superlinear (accelerating) spreading, but at a reduced rate. In contrast, pressure-limited growth results in linear (constant-speed) spreading in the long-time limit. We characterize how the expansion speed depends on the maximum growth rate, the limiting pressure value, and the effective fluid friction. Interestingly, nutrient-limited growth results in a phase transition: depending on the nutrient supply and how efficiently nutrient is converted to biomass, the spreading can be either superlinear or sublinear (decelerating). We predict the phase boundary in terms of these parameters and confirm with simulations. Thus, our results suggest that when an expansion slowdown is observed, its dominant cause is likely nutrient depletion. More generally, our work suggests that cell-level growth limitations can be inferred from population-level dynamics, and it offers a methodology for connecting these two scales.
\end{abstract}

\maketitle

\section*{Significance}
Packed microbial communities, such as colonies and biofilms, have diverse spreading dynamics, but the cell-level mechanisms that underlie these dynamics are not always clear. Here we use a fluid dynamics model to ask whether cell-level growth mechanisms can be inferred from community-level spreading dynamics. We find that different types of growth limitation lead to different spreading dynamics: timed growth arrest leads to exponential spread, pressure-limited growth leads to linear spread, and nutrient-limited growth leads to a phase transition between exponential and sublinear spread. Our work suggests that expansion slowdowns are predominantly due to nutrient depletion, and it offers a road map for quantitatively connecting the cell and population scales in proliferating communities.

\section*{Introduction} 
Bacterial communities such as biofilms are prevalent in nature. Biofilm formation provides protection against antibacterial environmental components such as antibiotics, extreme pH, and heavy metals \cite{Koo2017-uj,Luo2021-ve,Hathroubi2017-mo,McNeill2003-or,Teitzel2003-vd}. Biofilm formation also allows cells to attach to particular surfaces and promotes cell differentiation \cite{Flemming2016-xm}. Because of their ubiquity, researchers have long been interested in how biofilms form and what governs their growth dynamics.

At early stages, the number of cells in a biofilm, and hence the biofilm size, will grow exponentially \cite{Shao2017-je,Warren2019-um,Allen2019-rt}. However, this unbounded growth will not persist indefinitely due to unavoidable constraints. One well-studied constraint is nutrient access. Biofilms with more nutrients grow larger than those with fewer nutrients \cite{Allen2018-xb,Matsushita1990-uc,Siri2024-tm,Marsden2017-ak,Bottura2022-tt}. Nutrients can affect the total number of cells produced in the biofilm \cite{Allen2018-xb} as well as extracellular matrix production and biofilm structural properties \cite{Allen2018-xb,Siri2024-tm,Donlan2002-vq,Zhang2014-rj}. In addition to nutrient constraints, physical properties of the environment can affect the growth of biofilms. For example, biofilms of certain strains of bacteria can expand further on rougher surfaces \cite{Bottura2022-tt,Yan2017-pk,Seminara2012-yh}. Osmotic pressure is essential for expansion \cite{Yan2017-pk,Seminara2012-yh}, and yet higher physical pressure also leads to longer doubling time of cells \cite{Delarue2016-qp,Alric2022-aa,Kumar2013-as}. Finally, apart from external factors, bacteria may undergo internal processes that slow or arrest cell growth. A ubiquitous example is sporulation, where a cell transitions to a long-lived, non-growing state. Sporulation is a temporal process that can be triggered by unfavorable environmental conditions like nutrient depletion \cite{Khanna2020-bk,Tan2014-ya}, and is also linked to biofilm formation and matrix production \cite{Tan2014-ya,Hamon2001-xe,Srinivasan2018-il}. Despite extensive research on these limiting factors, a systematic understanding of their effects on the community-level growth dynamics is lacking.

Here, we use an active fluid model to explore the effects of various growth constraints on biofilm expansion dynamics. We separately consider three types of constraints: temporal growth arrest, pressure-limited growth, and nutrient-limited growth. Under the simplifying assumptions of radial symmetry and constant biofilm density, we find that each constraint leads to a distinct type of expansion dynamics. Temporal growth arrest produces superlinear (accelerating) expansion with a rate that decreases with the arrest rate; pressure-limited growth produces linear (constant-speed) expansion; and nutrient-limited growth produces a phase transition: expansion can be either super- or sublinear (decelerating) depending on the nutrient supply level and biomass conversion rate. We verify our analytic findings with simulations and confirm that our results hold in one, two, and three spatial dimensions. Our work connects cell-level growth mechanisms to population-level dynamics and suggests that nutrient depletion is the primary cause of expansion slowdown in bacterial communities.

\section*{Materials and Methods}
Active fluid models are widely used to investigate bacterial colony expansion and patterning \cite{Seminara2012-yh,Giometto2018-hn,alert2019active,Martinez-Calvo2025-yl,Ye2024-sg}. The general continuity equation for a multi-component active fluid model is
\begin{equation}
\label{eq:cte}
 \frac{\partial \rho_i}{\partial t}+\vec{\nabla} \cdot (\vec{v}\rho_i)= f_i(\rho_1,\rho_2,\dots),
\end{equation}
where $\rho_i(\vec{r},t)$ is the density of the $i$th component of the fluid that comprises the bacterial biofilm (e.g., growing or non-growing cells), $\vec{v}(\vec{r},t)$ is the velocity field of the fluid, and $f_i(\rho_1,\rho_2,\dots)$ is the active term that generates or removes mass of the $i$th component (e.g., due to cell growth or sporulation) which we will define for the three constraint types shortly. Here the position vector is $\vec{r} = r$, $(r,\theta)$, or $(r,\theta,\phi)$ in one, two, or three dimensions (we consider all three cases in this work), where $r$ is the radius, and $\theta$ and $\phi$ are the polar and azimuthal angles, respectively.

\begin{figure*}[hbt!]
\centering
\includegraphics[width=0.8\linewidth]{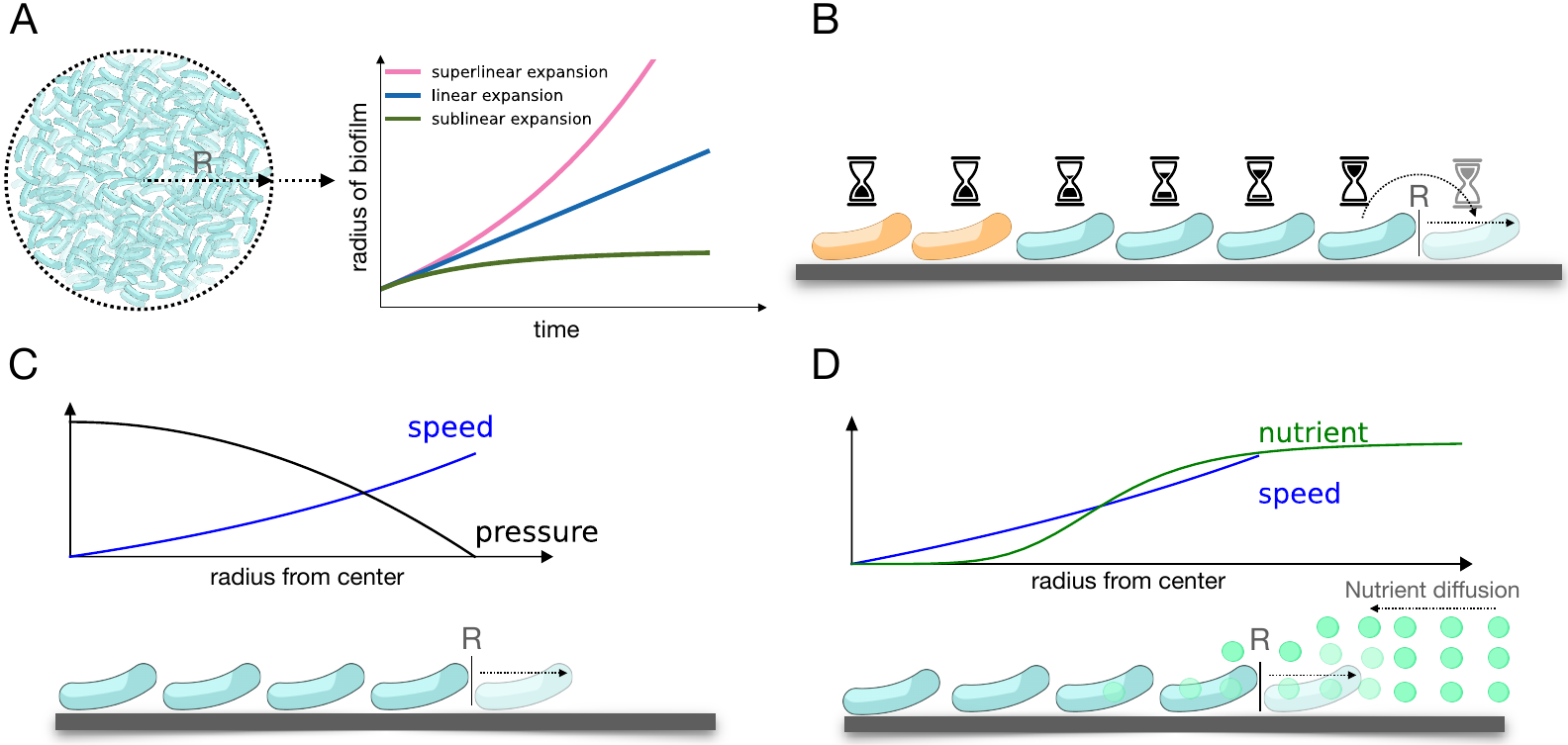}
\caption{Setup and three types of growth limitation. (A) We model a biofilm as a radially expanding active fluid and ask how the growth limitation type determines the expansion dynamics. (B) Type 1: temporal growth arrest, e.g.\ by sporulation at a fixed rate. (C) Type 2: Pressure-limited growth. Pressure is highest in the center and expansion speed is highest at the edge. (D) Type 3: Nutrient-limited growth. Diffusing nutrients are consumed by cells, leading to cell growth and biofilm expansion.}
\label{fig:generalidea}
\end{figure*}

Because biofilms are densely packed, we make the simplifying assumption that the density of cells is constant within the biofilm. Furthermore, because we are primarily interested in the radial expansion dynamics, we assume radial symmetry (Fig.\ \ref{fig:generalidea}A). Thus, Eq.\ \ref{eq:cte} depends only on the radial coordinate $r$. With these assumptions, we have
\begin{equation}
    \label{eq:rho}
    \sum_i\rho_i(r,t) = c\Theta[R(t)-r],
\end{equation}
where $c$ is the constant density of cells, $\Theta$ is the Heaviside function (equal to one for nonnegative argument and zero otherwise), and $R(t)$ is the radius of the biofilm.

Summing Eq.\ \ref{eq:cte} over $i$ and inserting Eq.\ \ref{eq:rho} gives, inside the biofilm,
\begin{equation}
\label{eq:dv}
    \frac{c}{r^{d-1}}\frac{\partial}{\partial r}(r^{d-1}v) = \sum_i f_i(\rho_1,\rho_2,\dots)
    \qquad {\rm for}\ r \le R(t),
\end{equation}
where the densities $\rho_i$ on the left have summed to the constant $c$, which vanishes under the time derivative and comes out of the spatial derivative. Further, radial symmetry has left only the radial part of the divergence operator in spatial dimension $d$, acting on the radial component $v(r,t)$ of the velocity vector. Eq.\ \ref{eq:dv} is subject to the boundary conditions
\begin{equation}
\label{eq:vbc}
    v(0,t) = 0, \qquad
    v(R,t) = \frac{dR}{dt},
\end{equation}
which state that the fluid has no velocity at the center (by symmetry) and that its velocity at the edge must match the growth of the biofilm radius.

We now describe the three types of growth constraints.

\subsection*{Temporal growth arrest}
The first type of growth limitation we consider is timed growth arrest (Fig.\ \ref{fig:generalidea}B). A prominent example of growth arrest is sporulation, where cells transition to a non-growing state \cite{Khanna2020-bk,Tan2014-ya}; therefore we will refer to growth arrest as sporulation in what follows. While sporulation can be coupled to environmental factors \cite{Khanna2020-bk, Tan2014-ya,Hamon2001-xe,Srinivasan2018-il}, for simplicity in this work we consider only internally timed, memoryless sporulation, and we discuss generalizations at the end. Specifically, we introduce a constant sporulation rate $k$ for every growing cell.

In this case, the fluid contains two components: a density $\rho_1(r,t)$ of growing cells and a density $\rho_2(r,t)$ of non-growing cells. Calling the growth rate $g$, the active terms are
\begin{equation}
\label{eq:f1}
    f_1 = g\rho_1 -k\rho_1, \qquad
    f_2 = k\rho_1,
\end{equation}
which represent, respectively, (1) the gain of growing cells due to growth and the loss of growing cells due to sporulation, and (2) the gain of spores due to the sporulation of growing cells.

\subsection*{Pressure-limited growth}
The second type of growth limitation that we consider is a reduction in growth rate due to pressure from surrounding cells (Fig.\ \ref{fig:generalidea}C). In a packed community, the growth of any cell requires pushing other cells out of the way. This will not be possible indefinitely if there is friction between cells (or with the substrate or other components of the biofilm). Research has shown that higher pressure leads to slower growth in microbial communities \cite{Delarue2016-qp,Alric2022-aa,Kumar2013-as}.

To implement pressure-limited growth, we consider a single-component fluid with constant density $\rho_1 = c$ and make its growth rate a decreasing function of pressure $p(r,t)$. The active term is then
\begin{equation}
\label{eq:f2}
    f = \frac{gcp_0}{p+p_0},
\end{equation}
where $g$ is the maximum growth rate, and $p_0$ is the pressure at which growth is half-maximal \cite{Ye2024-sg}. The pressure is then related to the friction coefficient $\xi$ via the velocity \cite{Martinez-Calvo2025-yl} as $\vec{\nabla} p=-\xi \vec{v}$, or
\begin{equation}
\label{eq:p}
    \frac{\partial p}{\partial r} = -\xi v
\end{equation}
with radial symmetry. The friction coefficient relates the pressure gradient that drives
expansion to the expansion velocity. The pressure obeys the boundary condition
\begin{equation}
\label{eq:pbc}
    p(R,t) = 0,
\end{equation}
reflecting the fact that pressure vanishes at the edge of the fluid.

\subsection*{Nutrient-limited growth}
The third type of growth limitation that we consider is a nutrient-dependent growth rate (Fig.\ \ref{fig:generalidea}D). Nutrient supply is a broadly investigated determinant of cell growth in biofilms \cite{Shao2017-je,Warren2019-um,Allen2019-rt,Allen2018-xb,Matsushita1990-uc,Siri2024-tm,Marsden2017-ak,Bottura2022-tt}. To implement nutrient-dependent growth, we again consider a single-component fluid with constant density $\rho_1 = c$. We make its growth rate a Monod function $gn/(n+n_g)$ of the nutrient concentration $n(r,t)$, where $g$ is the maximum growth rate, and $n_g$ is the nutrient concentration at which growth is half-maximal \cite{monod1949growth}. Because the bulk of cells will experience nutrient concentrations well below $n_g$, for simplicity we consider this function in its linear regime, giving
\begin{equation}
\label{eq:f3}
f = \frac{gc n}{n_g}
\end{equation}
for the active term.

The nutrient concentration obeys a diffusion equation with coefficient $D$,
\begin{equation}
\label{eq:ndgnutrient_edited}
\frac{\partial n}{\partial t}=D\nabla ^2n-\gamma\frac{gcn}{n_g},
\end{equation}
where the last term reflects consumption by the cells, and $\gamma$ accounts for the conversion of nutrient into cell biomass. Larger $\gamma$ means that it requires more nutrient to create one cell. With radial symmetry, Eq.\ \ref{eq:ndgnutrient_edited} becomes
\begin{equation}
\label{eq:ndgnutrient_edited2}
\frac{\partial n}{\partial t}=\frac{D}{r^{d-1}}\frac{\partial}{\partial r}\left(r^{d-1}\frac{\partial n}{\partial r}\right)-\gamma\frac{gcn}{n_g}
\end{equation}
for spatial dimension $d$.

The nutrient concentration obeys the initial condition
\begin{equation}
\label{eq:nic}
    n(r,0) = n_0 \qquad {\rm for}\ r\leq L,
\end{equation}
where $L$ is the system radius. We assume reflective boundaries for the nutrient at $r=L$, meaning that a fixed amount of nutrient is provided initially and not replenished. The biofilm grows until it reaches the edge, $R=L$, or until all nutrient is consumed. This protocol is consistent with many experimental setups \cite{Matsushita1990-uc,Seminara2012-yh,Srinivasan2018-il}.

\subsection*{Numerical solution methods}
We solve our model equations numerically by discretizing time and space using the standard Euler method. For temporal sporulation, we simultaneously solve Eq.\ \ref{eq:cte} for $\rho_1(r,t)$ and $\rho_2(r,t)$, and Eq.\ \ref{eq:dv} for $v(r,t)$. For pressure-limited growth, we simultaneously solve Eq.\ \ref{eq:dv} for $v(r,t)$ and Eq.\ \ref{eq:p} for $p(r,t)$. For nutrient-limited growth, we simultaneously solve Eq.\ \ref{eq:dv} for $v(r,t)$ and Eq.\ \ref{eq:ndgnutrient_edited2} for $n(r,t)$. In all cases, we update $R(t)$ at each time step using Eq.\ \ref{eq:vbc}. See Appendix A for details and \cite{code} for the code.

\section*{Results}
Here we consider results for the case of two spatial dimensions ($d=2$). We find qualitatively similar results in one and three dimensions, which are treated in Appendix B.

\subsection*{Temporal growth arrest produces superlinear spread}
With the active terms in Eq.\ \ref{eq:f1} for temporal sporulation, Eq.\ \ref{eq:dv} with $d=2$ reads
\begin{equation}
\label{eq:v_eq1}
    \frac{\partial}{\partial r}(vr)=\frac{g\rho_1r}{c}.
\end{equation}
Integrating both sides over $r$ from $0$ to $R(t)$ gives
\begin{equation}
\label{eq:Rv}
    v(R,t)R = \frac{gC_1}{2\pi c},
\end{equation}
where
\begin{equation}
\label{eq:C1}
    C_1 = 2\pi\int_0^{R}\rho_1rdr
\end{equation}
is the total amount of growing cells. Recalling from Eq.\ \ref{eq:vbc} that $v(R,t) = dR/dt$ and noting that $(dR/dt)R = (1/2)d(R^2)/dt$, Eq.\ \ref{eq:Rv} becomes
\begin{equation}
\frac{1}{2}\frac{d(R^2)}{dt} = \frac{gC_1}{2\pi c},
\end{equation}
or, upon integrating,
\begin{equation}
\label{eq:Rint}
    R(t) = \left[R_0^2+\frac{g}{\pi c}\int_0^tC_1(t')dt'\right]^{1/2},
\end{equation}
where $R_0$ is the initial biofilm radius. This result makes clear how the expansion $R(t)$ is driven by the total amount of growing cells $C_1$.

To find $C_1(t)$, we differentiate Eq.\ \ref{eq:C1} with respect to time. Because both $R$ and $\rho_1$ are time-dependent, we get two terms,
\begin{equation}
\label{eq:dC1}
    \frac{1}{2\pi}\frac{dC_1}{dt}
    = \rho_1(R,t)R\frac{dR}{dt} + \int_0^R\frac{\partial\rho_1}{\partial t}rdr.
\end{equation}
The time derivative of $\rho_1$ is then given by the first component of Eq.\ \ref{eq:cte} with the active terms of Eq.\ \ref{eq:f1},
\begin{equation}
\label{eq:drho1}
    \frac{\partial\rho_1}{\partial t} = -\frac{1}{r}\frac{\partial}{\partial r}(v\rho_1r) + g\rho_1 - k\rho_1.
\end{equation}
Inserting Eq.\ \ref{eq:drho1} into Eq.\ \ref{eq:dC1} obtains
\begin{equation}
    \frac{1}{2\pi}\frac{dC_1}{dt}
    = \rho_1(R,t)R\frac{dR}{dt} - v(R,t)\rho_1(R,t)R + (g-k)\frac{C_1}{2\pi}.
\end{equation}
Again recognizing that $v(R,t) = dR/dt$, the first two terms on the righthand side cancel, and simple time integration yields
\begin{equation}
\label{eq:C1t}
    C_1(t) = C_1(0)e^{(g-k)t},
\end{equation}
where $C_1(0)$ is the initial amount of growing cells.
Inserting Eq.\ \ref{eq:C1t} into Eq.\ \ref{eq:Rint} and integrating obtains our main result for this case, the radial expansion over time:
\begin{equation}
\label{eq:R1}
    R(t) = \left\{R_0^2+\frac{C_1(0)}{\pi c}\frac{g}{g-k}\left[e^{(g-k)t}-1\right]\right\}^{1/2}.
\end{equation}
Eq.\ \ref{eq:R1} is initialized by specifying $R_0$ and $\rho_1(r,0)$, which gives $C_1(0)$ via Eq.\ \ref{eq:C1}.

\begin{figure*}[hbt!]
\centering
\includegraphics[width=0.8\linewidth]{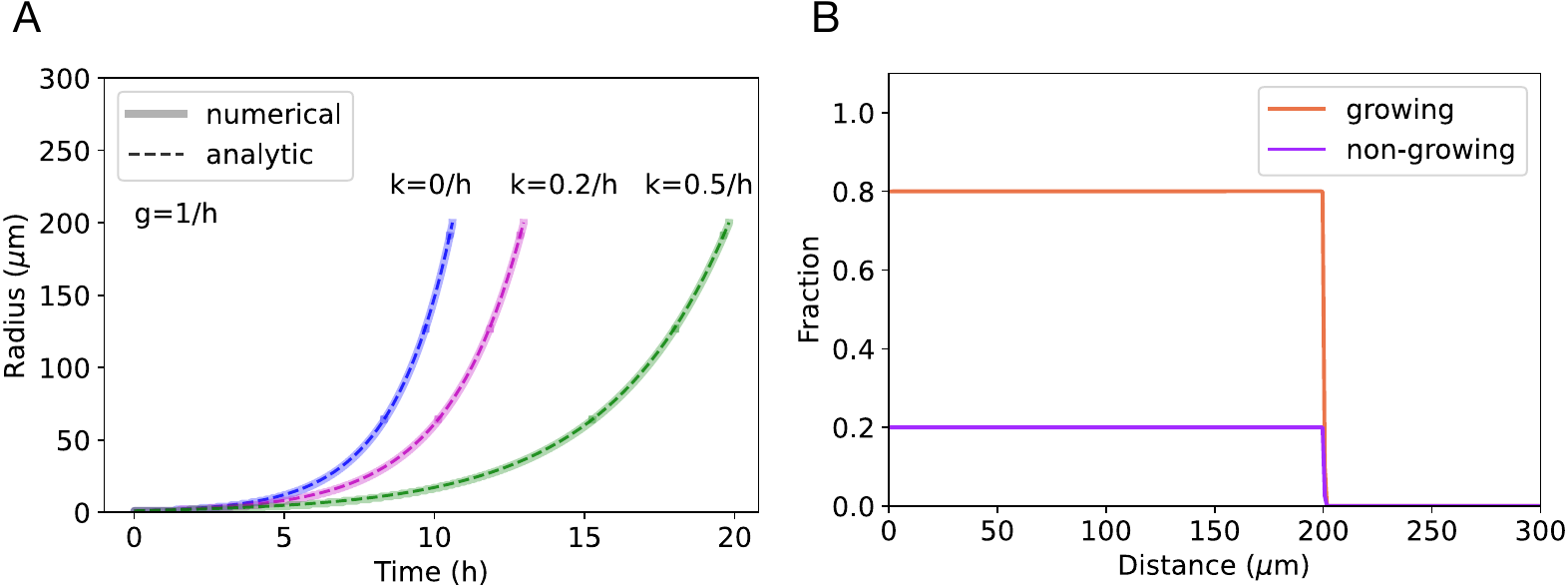}
\caption{Temporal growth arrest. (A) Radial expansion for different sporulation rates $k$, from numerical solution and analytic expression (Eq.\ \ref{eq:R1}). We see that growth arrest from sporulation slows expansion, but expansion remains superlinear in time. (B) Snapshot of density fractions $\rho_1(r,t)/c$ (orange) and $\rho_2(r,t)/c$ (purple) over distance $r$ with $k = 0.2$/h. We see that the densities are uniform throughout the biofilm. Parameters are $g=1$/h and $c=1$/$\mu$m$^2$, and the biofilm starts at size $R_0=1$ $\mu$m with all growing cells, i.e., $\rho_1(r,0) = c\Theta(R_0-r)$.}
\label{fig:2dtimer simulation result}
\end{figure*}

We see from Eq.\ \ref{eq:R1} that in order to expand, growth must outpace sporulation, $g>k$. Further, we see that at sufficiently long times we may neglect $R_0$ and the $-1$, such that
\begin{equation}
    R(t) \sim e^{(g-k)t/2}.
\end{equation}
Thus, sporulation slows the expansion rate but does not prevent the biofilm from spreading exponentially. Indeed, Eq.\ \ref{eq:R1} is shown in Fig.\ \ref{fig:2dtimer simulation result}A, and we see that the expansion is superlinear.

Figure \ref{fig:2dtimer simulation result}B shows a simulation snapshot of the radial profile of the biofilm, meaning that it plots $\rho_1(r,t)$ (orange) and $\rho_2(r,t)$ (purple) at a particular $t$. We see that the densities of growing cells and spores are uniform in space. This is because, in this model, both growth and sporulation are memoryless processes, meaning when (and where) a cell sporulates is independent of when (and where) it was born.
The observation of uniformity allows us to calculate the density fractions analytically. Specifically, the fact that $\rho_1$ does not depend on $r$ takes Eq.\ \ref{eq:drho1} to
\begin{equation}
    \frac{\partial\rho_1}{\partial t} = -\frac{\rho_1}{r}\frac{\partial}{\partial r}(vr) + g\rho_1 - k\rho_1.
\end{equation}
Inserting Eq.\ \ref{eq:v_eq1} obtains
\begin{equation}
    \frac{\partial\rho_1}{\partial t} = -\frac{g}{c}\rho_1^2 + g\rho_1 - k\rho_1.
\end{equation}
This ordinary differential equation has fixed points at $\rho_1/c = 0$ and $1-k/g$, with only the latter stable. Thus, the fractions approach
\begin{equation}
    \frac{\rho_1}{c} = 1-\frac{k}{g}, \qquad \frac{\rho_2}{c} = \frac{k}{g}
\end{equation}
at long times. We see that, as expected, the fractions of growing and non-growing cells decrease and increase, respectively, with the sporulation rate $k$ (scaled by the growth rate $g$).

\subsection*{Pressure-limited growth produces linear spread}

With the active term in Eq.\ \ref{eq:f2} for pressure-limited growth, Eq.\ \ref{eq:dv} with $d=2$ reads
\begin{equation}
\label{eq:v_eq2}
    \frac{1}{r}\frac{\partial}{\partial r}(rv)=\frac{gp_0}{p+p_0}.
\end{equation}
Differentiating both sides with respect to $r$ obtains
\begin{equation}
\label{eq:v2order_edited}
    \frac{\partial^2 v}{\partial r^2} + \frac{1}{r}\frac{\partial v}{\partial r} - \frac{v}{r^2} = -\frac{g}{p_0}\frac{\partial p}{\partial r}\left[\left(1+\frac{p}{p_0}\right)^2\right]^{-1}.
\end{equation}
Because growth will be driven by the leading edge of the biofilm where pressure is low, we assume $p/p_0\ll 1$ and neglect this term. Then, recalling from Eq.\ \ref{eq:p} that $\partial p/\partial r = -\xi v$, Eq.\ \ref{eq:v2order_edited} becomes
\begin{equation}
\label{eq:bessel}
    x^2\frac{\partial^2 v}{\partial x^2} + x\frac{\partial v}{\partial x} -(x^2+1)v = 0,
\end{equation}
where we have defined $x=r/\lambda$ and $\lambda = \sqrt{p_0/(g\xi)}$. Equation \ref{eq:bessel} is solved by
\begin{equation}
\label{eq:vAB}
    v(x,t) = A(t)I_1(x) + B(t)K_1(x),
\end{equation}
where $I_1(x)$ and $K_1(x)$ are modified Bessel functions of order one. Because $K_1(x)$ diverges at $x=0$, the boundary condition $v(0,t) = 0$ (Eq.\ \ref{eq:vbc}) implies $B=0$. We find $A$ using the second boundary condition $p(R,t) = 0$ (Eq.\ \ref{eq:pbc}).
Specifically, evaluating Eq.\ \ref{eq:v_eq2} at $r=R$, where $p=0$, obtains
\begin{equation}
\label{pbc}
    \frac{v(R,t)}{R} + \left.\frac{\partial v}{\partial r}\right|_R = g.
\end{equation}
Inserting Eq.\ \ref{eq:vAB} (with $B=0$) and solving for $A$ obtains
\begin{equation}
    A(t) = \frac{g\lambda R}{\lambda I_1(R/\lambda) + RI'_1(R/\lambda)},
\end{equation}
where $I_1'(x) = dI_1/dx$. Thus, evaluating Eq.\ \ref{eq:vAB} at $r=R$, and recalling from Eq.\ \ref{eq:vbc} that $v(R,t) = dR/dt$, we have
\begin{equation}
\label{eq:Rode2}
    \frac{dR}{dt} = \frac{g\lambda RI_1(R/\lambda)}{\lambda I_1(R/\lambda) + RI'_1(R/\lambda)},
\end{equation}
which is an ordinary differential equation for our quantity of interest, the radial expansion over time, $R(t)$.

Equation \ref{eq:Rode2} is not generally solvable. However, for large argument the Bessel function limits to $I_1(x) \to e^x/\sqrt{2\pi x}$ and therefore $I'_1(x) \to I_1(x)[1-1/(2x)] \approx I_1(x)$. Thus, for $R\gg \lambda$, the Bessel functions in Eq.\ \ref{eq:Rode2} cancel, and it becomes $dR/dt \to g\lambda R/(\lambda + R) \approx g\lambda$. Recalling that $\lambda = \sqrt{p_0/(g\xi)}$, we obtain our main result for this case,
\begin{equation}
\label{eq:linear_speed}
    \frac{dR}{dt} \to \sqrt{\frac{p_0g}{\xi}}
    \qquad {\rm for}\ R \gg \sqrt{\frac{p_0}{g\xi}}.
\end{equation}
This result states that at long times, the spreading speed $dR/dt$ becomes a constant. That is, the radius expands linearly over time.

\begin{figure*}[hbt!]
\centering
\includegraphics[width=0.8\linewidth]{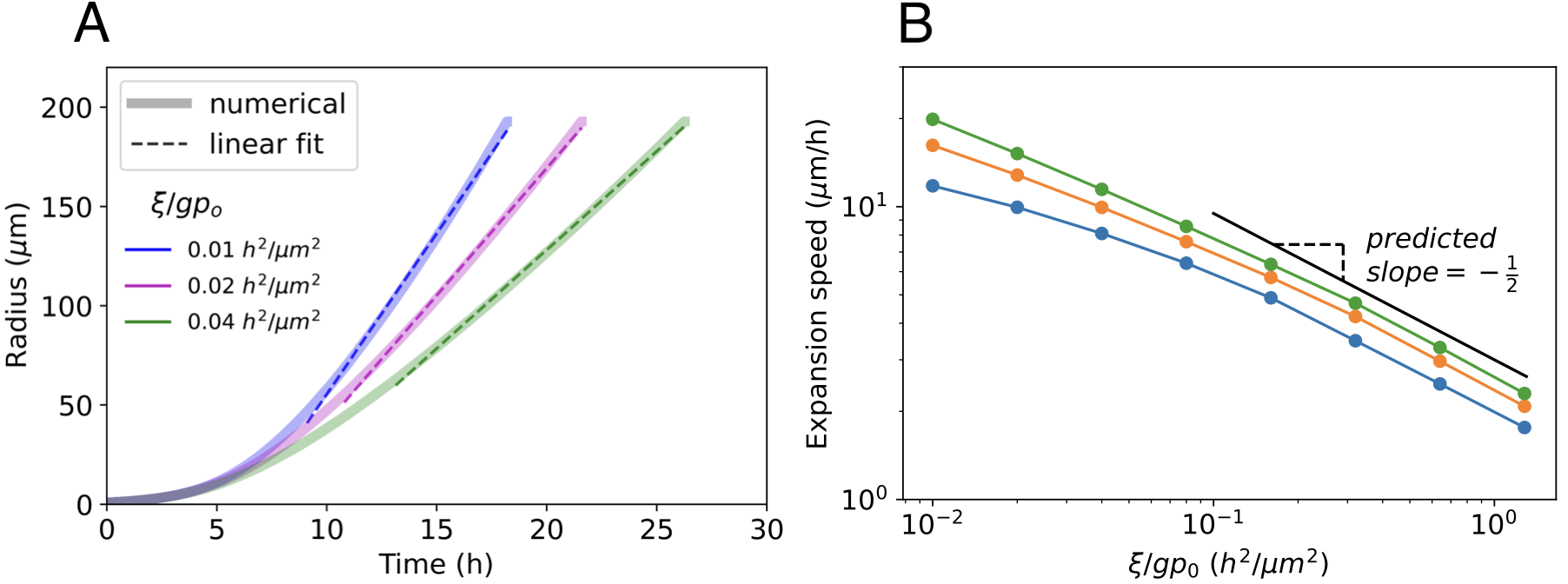}
\caption{Pressure-limited growth. (A) Radial expansion for different parameters (see legend). We see that expansion is linear in time at long times. (B) Fitted speed vs.\ $\xi/gp_0$. Black line shows predicted dependence from Eq.\ \ref{eq:linear_speed}. Parameters are $c=1$/$\mu$m$^2$ and $R_0=1$ $\mu$m. In A, $g=1$/h. In B, $\xi$ is varied for several choices of $g$ and $p_0$.}
\label{fig:2dpressureresult}
\end{figure*}

We test this result against numerics in Fig.\ \ref{fig:2dpressureresult}A, where we see that indeed, $R(t)$ becomes linear at long times.
In Fig.\ \ref{fig:2dpressureresult}B we plot the fitted speed against the quantity $\xi/(gp_0)$ on a log-log scale, and we see that the slope approaches the value of $-1/2$ predicted by Eq.\ \ref{eq:linear_speed}.

\subsection*{Nutrient-limited growth produces a phase transition}

With the active term in Eq.\ \ref{eq:f3} for nutrient-limited growth, Eq.\ \ref{eq:dv} with $d=2$ reads
\begin{equation}
\label{eq:v_eq3}
    \frac{\partial}{\partial r}(vr)=\frac{gnr}{n_g}.
\end{equation}
We begin our analysis of this equation using a similar approach to that for the temporal sporulation case.
Integrating both sides over $r$ from $0$ to $R(t)$ gives
\begin{equation}
\label{eq:Rv3}
    v(R,t)R = \frac{gN}{2\pi n_g},
\end{equation}
where
\begin{equation}
\label{eq:N}
    N = 2\pi\int_0^{R}nrdr
\end{equation}
is the total amount of nutrient within the biofilm. Recalling from Eq.\ \ref{eq:vbc} that $v(R,t) = dR/dt$ and noting that $(dR/dt)R = (1/2)d(R^2)/dt$, Eq.\ \ref{eq:Rv3} becomes
\begin{equation}
\label{dR2}
\frac{1}{2}\frac{d(R^2)}{dt} = \frac{gN}{2\pi n_g},
\end{equation}
or, upon integrating,
\begin{equation}
\label{eq:Rint3}
    R(t) = \left[R_0^2+\frac{g}{\pi n_g}\int_0^tN(t')dt'\right]^{1/2},
\end{equation}
where $R_0$ is the initial biofilm radius. We see that expansion is driven by the total amount of accessible nutrient $N(t)$.

To find $N(t)$, we differentiate Eq.\ \ref{eq:N} with respect to time. As before, because both $R$ and $n$ are time-dependent, we get two terms,
\begin{equation}
\label{eq:dN1}
    \frac{1}{2\pi}\frac{dN}{dt}
    = n(R,t)R\frac{dR}{dt} + \int_0^R\frac{\partial n}{\partial t}rdr.
\end{equation}
Because $v(R,t) = dR/dt$, the quantity $R(dR/dt)$ in the first term is given by Eq.\ \ref{eq:Rv3}. Meanwhile, $\partial n/\partial t$ in the second term is given by the nutrient dynamics in Eq.\ \ref{eq:ndgnutrient_edited2} with $d=2$. Inserting both of these obtains
\begin{equation}
\label{eq:dN2_edited}
    \frac{1}{2\pi}\frac{dN}{dt}
    = n(R,t)\frac{gN}{2\pi n_g} + \int_0^R\left[\frac{D}{r}\frac{\partial}{\partial r}\left(r\frac{\partial n}{\partial r}\right)-\gamma\frac{gcn}{n_g}\right]rdr.
\end{equation}
The first term in brackets is the integral of a derivative and thus gets evaluated at the boundaries, while the second term is proportional to $N$ via Eq.\ \ref{eq:N}. Thus, Eq.\ \ref{eq:dN2_edited} becomes
\begin{equation}
\label{eq:dN3_edited}
    \frac{dN}{dt}
    = \frac{g}{n_g}[n(R,t)-\gamma c]N + 2\pi DR\left.\frac{\partial n}{\partial r}\right|_R.
\end{equation}
We see that the dynamics of $N$ depend not only on the value of the nutrient concentration at the boundary but, due to diffusion, also its slope there.

Equation \ref{eq:dN3_edited} is difficult to solve in general, and so we make progress in two limits: $D=0$ and $D\to\infty$. We will see that both limits give the same qualitative result---a phase transition between superlinear and sublinear spread---but with different phase boundaries. Furthermore, we will see that the numerical solution for any $D$ interpolates between these two boundaries, and we will provide the parametric criterion for when one or the other limit holds.

\subsubsection*{Slow-diffusion limit}

For $D=0$, there is no diffusion of the nutrient into the biofilm. Cells simply consume nutrient as they spread. Therefore, the nutrient concentration at the boundary $n(R,t)$ remains equal at all times to the initial nutrient concentration $n_0$. Thus, Eq.\ \ref{eq:dN3_edited} with $D=0$ reads
\begin{equation}
\label{eq:dN4}
    \frac{dN}{dt}
    = \frac{g}{n_g}(n_0-\gamma c)N.
\end{equation}
Integrating gives
\begin{equation}
\label{eq:Nt}
    N(t) = N(0)e^{g(n_0-\gamma c)t/n_g},
\end{equation}
where $N(0) = \pi R_0^2n_0$ by Eqs.\ \ref{eq:nic} and \ref{eq:N}.
Inserting Eq.\ \ref{eq:Nt} into Eq.\ \ref{eq:Rint3} and integrating obtains the radial expansion over time,
\begin{equation}
\label{eq:R3}
    \frac{R(t)}{R_0} = \left\{1+\frac{n_0}{n_0-\gamma c}\left[e^{g(n_0-\gamma c)t/n_g}-1\right]\right\}^{1/2}.
\end{equation}
The qualitative behavior of Eq.\ \ref{eq:R3} depends critically on the sign of $n_0-\gamma c$. At long times, the radius either grows exponentially or saturates to a constant value,
\begin{equation}
\label{eq:phase}
    \frac{R(t)}{R_0} \to
    \begin{cases}
        \sim e^{g(n_0-\gamma c)t/(2n_g)} & n_0>\gamma c \\
        \sqrt{\gamma c/(\gamma c-n_0)} & n_0<\gamma c.
    \end{cases}
\end{equation}
Thus, for $D=0$, the boundary $n_0=\gamma c$ separates the dynamics into two phases: superlinear spread ($n_0>\gamma c$) and sublinear spread ($n_0<\gamma c$).

The reason that the dynamics depend strongly on the sign of $n_0 -\gamma c$ is that $\gamma c$ is the minimal nutrient concentration needed to sustain a cell density $c$ with metabolic conversion parameter $\gamma$. Therefore, if the external nutrient concentration $n_0$ is greater than this minimum, cells will always be able to sustain exponential growth by advancing to new territory, and the biofilm will spread exponentially. However, if $n_0$ is less than this minimum, the amount of nutrients acquired by spreading will be insufficient to sustain that spread, and the biofilm will slow down.

\subsubsection*{Fast-diffusion limit}

For $D\to\infty$, diffusion is so large that the nutrient profile is always uniform, $n(r,t) \to n(t)$. Equivalently, $dn/dr \to 0$. The last term of \ref{eq:dN3_edited} becomes the product of term that diverges, $D$, and a term that vanishes, $(dn/dr)|_R$. We assume that $(dn/dr)|_R$ vanishes more strongly than $D$ diverges, and this assumption will be confirmed post hoc when we test our predictions against our numerics. With this assumption, Eq.\ \ref{eq:dN3_edited} becomes
\begin{equation}
\label{eq:dN3_inf}
    \frac{dN}{dt} = \frac{g}{n_g}(n-\gamma c)N.
\end{equation}
Because it is uniform, the nutrient concentration $n$ is equal to the initial nutrient amount, $\pi L^2 n_0$, minus the amount consumed by the biofilm, divided by the system area, $\pi L^2$. The amount consumed by the biofilm is equal to its added area, $\pi R^2 - \pi R_0^2$, times the cell density $c$, times the metabolic conversion parameter $\gamma$. Thus, we have
\begin{equation}
\label{eq:nu_inf_n2_new}
    n =\frac{\pi L^2n_0-\gamma c(\pi R^2-\pi R_0^2)}{\pi L^2}.
\end{equation}
Due to its uniformity, $n$ is also equal to the nutrient concentration inside the biofilm,
\begin{equation}
\label{NRD}
    n=\frac{N}{\pi R^2}.
\end{equation}
Combining Eqs.\ \ref{eq:nu_inf_n2_new} and \ref{NRD} relates $N$ to $R$:
\begin{equation}
\label{NR}
N = \pi R^2[n_0-\gamma c(R^2-R_0^2)/L^2].
\end{equation}
In principle, Eqs.\ \ref{eq:nu_inf_n2_new} and \ref{NR} can be inserted into Eq.\ \ref{eq:dN3_inf} to give a separable first-order differential equation for $R$. However, it is more revealing to assess the superlinearity or sublinearity of $R(t)$ directly by calculating its acceleration, $a = d^2R/dt^2$.

To find the acceleration, we differentiate Eq.\ \ref{dR2} with respect to time,
\begin{equation}
\label{d2R}
\left(\frac{dR}{dt}\right)^2+R\left(\frac{d^2R}{dt^2}\right) = \frac{g}{2\pi n_g}\frac{dN}{dt},
\end{equation}
where the lefthand side comes from the chain rule. Inserting Eq.\ \ref{eq:Rv3} for $dR/dt = v(R,t)$, inserting Eq.\ \ref{eq:dN3_inf} for $dN/dt$, and solving for $a = d^2R/dt^2$ obtains
\begin{equation}
\label{a1}
a = \frac{g^2N}{2\pi n_g^2R}\left(n-\gamma c - \frac{N}{2\pi R^2}\right).
\end{equation}
Then, using Eq.\ \ref{NRD} for the last term and simplifying obtains
\begin{equation}
a = \frac{g^2N}{2\pi n_g^2R}\left(\frac{n}{2}-\gamma c\right).
\end{equation}
Finally, inserting Eq.\ \ref{eq:nu_inf_n2_new} for $n$ and simplifying obtains
\begin{equation}
\label{acc}
a = \frac{g^2N}{2\pi n_g^2R}\left[\frac{n_0}{2}-\frac{\gamma c}{2}\left(\frac{R^2-R_0^2}{L^2}\right)-\gamma c\right].
\end{equation}
At long times, given sufficient nutrient, the biofilm will approach the system edge, $R\to L$. If the initial biofilm size is much smaller than the system size, $R_0\ll L$, then Eq.\ \ref{acc} will approach
\begin{equation}
\label{acc2}
a \to \frac{g^2N}{4\pi n_g^2R}(n_0-3\gamma c)
\end{equation}
at long times.

Superlinear spread ($a>0$) or sublinear spread ($a<0$) is determined by the sign of the term in parentheses in Eq.\ \ref{acc2}. Thus, for $D\to\infty$, we see that the boundary $n_0=3\gamma c$ separates the dynamics into two phases: superlinear spread ($n_0>3\gamma c$) and sublinear spread ($n_0<3\gamma c$). This is the same qualitative behavior as the $D=0$ case, but the phase boundary has acquired an extra factor of 3. This means that within our model, a biofilm requires three times as much fast-diffusing nutrient as slow-diffusing nutrient to sustain superlinear spread. The reason is that fast diffusion allows more nutrient to be consumed at early times, leaving less nutrient at late times, which promotes a growth slowdown at late times.

\subsubsection*{Parametric criterion and numerical results}
What determines whether the system is closer to the slow- or fast-diffusion limit? We expect the key factor to be whether the nutrient can diffuse across the system lengthscale $L$ in the time it takes for the biofilm to grow. The growth timescale is order $1/g$, and the distance that the nutrient diffuses over this timescale is roughly $\sqrt{D/g}$. Therefore, we expect the slow-diffusion limit to hold if $\sqrt{D/g}$ is much less than $ L$, and we expect the fast-diffusion limit to hold if $\sqrt{D/g}$ is much greater than $ L$:
\begin{align}
{\rm Slow\ diffusion:\ } &\sqrt{D/g} \ll L, \nonumber\\
\label{regimes}
{\rm Fast\ diffusion:\ } &\sqrt{D/g} \gg L.
\end{align}
We now test the above results, as well as their regimes of applicability in Eq.\ \ref{regimes}, using our numerics. 

\begin{figure*}[hbt!]
\centering
\includegraphics[width=0.8\linewidth]{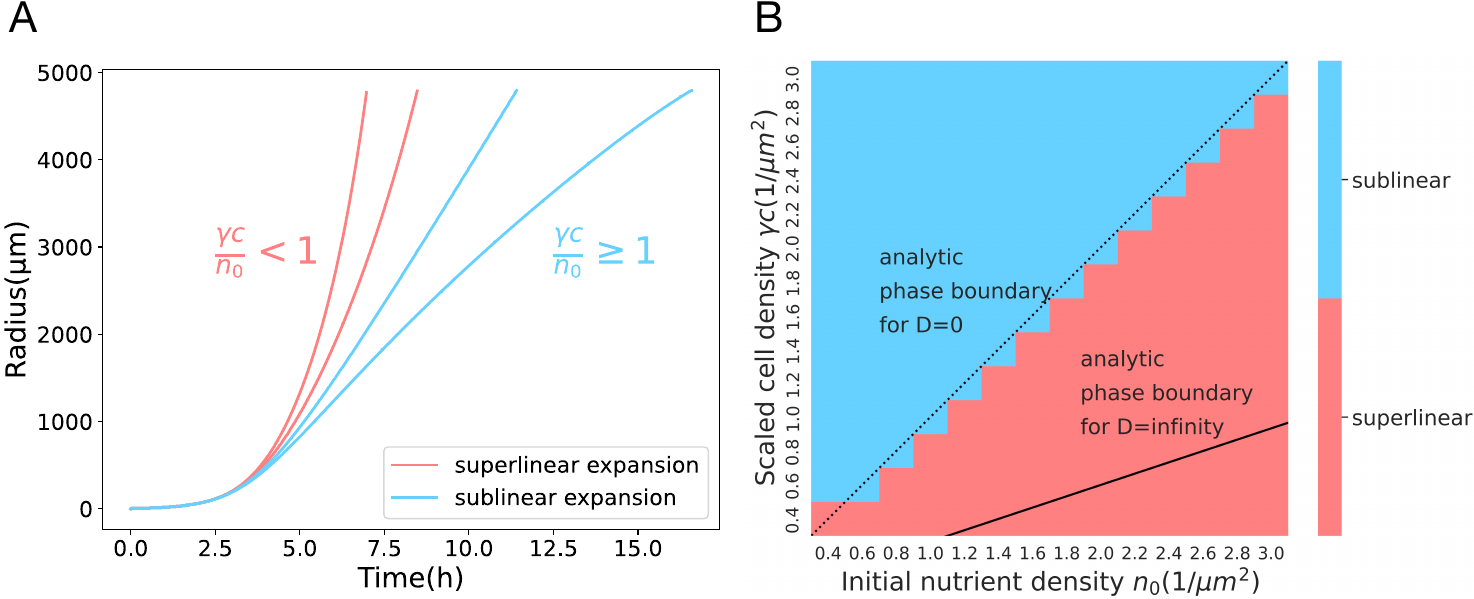}
\caption{Nutrient-limited growth with a large system size $L\gg\sqrt{D/g}$. (A) Radial expansion from numerical solution for different values of metabolic conversion parameter $\gamma$ (here $n_0=2.4$/$\mu$m$^2$ and $c=1$/$\mu$m$^2$). Color indicates whether curve is superlinear or sublinear at long times (last $0.1$ s of curve).
(B) Phase diagram for many values of $\gamma$ and $n_0$ (here $c=1$/$\mu$m$^2$), along with predicted phase boundaries $n_0 = \gamma c$ (dashed, Eq.\ \ref{eq:phase}) and $n_0 = 3\gamma c$ (solid, Eq.\ \ref{acc2}) in the slow- and fast-diffusion regimes, respectively. We see agreement with the slow-diffusion boundary, as expected for the large system size.Other parameters are $L = 6000$ $\mu$m, $g=1$/h, $D = 100$ $\mu$m$^2$/s, $n_g=1$ $\mu$M, and $R_0 = 6$ $\mu$m.}
\label{fig:2dnutrientsimulation_large_l}
\end{figure*}

\begin{figure*}[hbt!]
\centering
\includegraphics[width=0.8\linewidth]{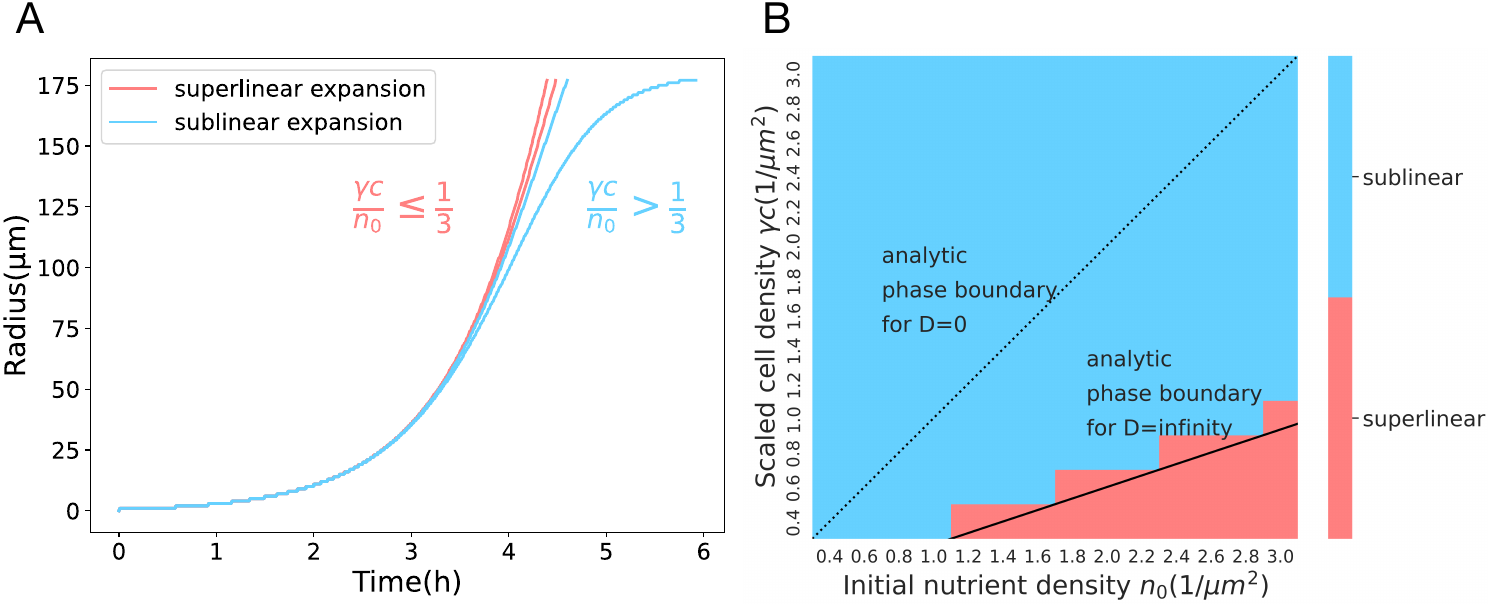}
\caption{Nutrient-limited growth with a small system size $L\ll\sqrt{D/g}$. (A) Radial expansion from numerical solution for different values of metabolic conversion parameter $\gamma$ (here $n_0=2.4$/$\mu$m$^2$ and $c=1$/$\mu$m$^2$). Color indicates whether curve is superlinear or sublinear at long times (last $0.1$ s of curve).
(B) Phase diagram for many values of $\gamma$ and $n_0$ (here $c=1$/$\mu$m$^2$), along with predicted phase boundaries $n_0 = \gamma c$ (dashed, Eq.\ \ref{eq:phase}) and $n_0 = 3\gamma c$ (solid, Eq.\ \ref{acc2}) in the slow- and fast-diffusion regimes, respectively. We see agreement with the fast-diffusion boundary, as expected for the small system size.
Other parameters are $L = 180$ $\mu$m, $g=1$/h, $D = 100$ $\mu$m$^2$/s, $n_g=1$ $\mu$M, and $R_0 = 1$ $\mu$m.}
\label{fig:2dnutrientsimulation_small_l}
\end{figure*}

To probe the regimes in Eq.\ \ref{regimes}, we choose typical values for the growth rate, $g = 1$/h, and diffusion constant, $D = 100$ $\mu$m$^2$/s, and we vary the system radius $L$. These values for $g$ and $D$ give $\sqrt{D/g} = 600$ $\mu$m. Figure \ref{fig:2dnutrientsimulation_large_l} shows results for a system size much larger than this value ($L=6000$ $\mu$m), where we expect the slow-diffusion limit to hold, and Fig.\ \ref{fig:2dnutrientsimulation_small_l} shows results for a system size much smaller than this value ($L=180$ $\mu$m), where we expect the fast-diffusion limit to hold.

We see in Fig.\ \ref{fig:2dnutrientsimulation_large_l}A that at long times, for large initial nutrient concentration (red curves) the biofilm spreads superlinearly, whereas for small initial nutrient concentration (blue curves) the biofilm spreads sublinearly. We see in Fig.\ \ref{fig:2dnutrientsimulation_large_l}B that the transition between these regimes is correctly predicted by the slow-diffusion limit (dashed line), as expected.

Similarly, we see in Fig.\ \ref{fig:2dnutrientsimulation_small_l}A that at long times, for large initial nutrient concentration (red curves) the biofilm spreads superlinearly, whereas for small initial nutrient concentration (blue curves) the biofilm spreads sublinearly. And we see in Fig.\ \ref{fig:2dnutrientsimulation_small_l}B that the transition between these regimes is correctly predicted by the fast-diffusion limit (solid line), as expected.

These numerical results confirm that nutrient-limited growth robustly leads to a phase transition between superlinear and sublinear spread, they validate the analytic predictions for the phase boundary in the slow- and fast-diffusion limits, and they illustrate the parameter regimes in which these limits apply.

\section*{Discussion}
Biofilm expansion is a fundamental process in active matter physics. Expansion does not persist indefinitely, and here we have recognized that the slowdown dynamics can reveal a lot about the expansion mechanism itself. Specifically, we have used an active fluid model to investigate the dynamics of slowdown due to three mechanisms: temporal growth arrest, pressure-limited growth, and nutrient-limited growth. Notably, we have found that each limiting mechanism slows down otherwise exponential expansion dynamics in distinct ways: temporal growth arrest leaves the expansion superlinear but slows the rate, pressure-limited growth slows the expansion to linear, and nutrient-limited leads to a phase transition between either superlinear or sublinear expansion. We have characterized how these dynamics depend on the physical parameters of the system and verified our analytic predictions with numerical results.

Our results connect expansion dynamics at the population scale with growth-limitation mechanisms at the cell scale, and they suggest that the latter could be inferred from the former. For example, we have found that neither temporal growth arrest nor pressure-limited growth alone can lead to deceleration of radial expansion, but nutrient-limited growth can. Therefore, when deceleration is observed, it is likely that nutrient limitation is the dominant cause. Alternatively, if temporal growth arrest or pressure limitation plays an role, it is likely coupled to nutrient limitation.

In the specific case of sporulation, all three limiting mechanisms likely play a role. For example, {\it Bacillus subtilis} sporulation follows a temporal program once induced \cite{Gauvry2019,Piggot2004}, and it is known to be triggered by both nutrient limitation \cite{Khanna2020-bk,Tan2014-ya,Grossman1988,Sonenshein2000,Lazazzera2000} and high cell density \cite{Grossman1988,Sonenshein2000,Lazazzera2000}. Our results suggest that the coupling to nutrient limitation is indispensable: if the density sensing were equivalent to pressure sensing, our finding that neither temporal arrest nor pressure limitation alone can decelerate expansion suggests that their combination may not either.

A natural direction for future work is to incorporate the aforementioned coupling among limitation mechanisms into the mathematical modeling, and thereby more rigorously investigate their combined effect on the expansion dynamics. One interesting consequence of coupling a multi-component mechanism like sporulation to other limitation mechanisms is that patterns are likely to emerge in the components. For example, both pressure and nutrient concentration are functions of radius and if coupled to sporulation could lead to radial patterns of spores. Expansion would then lead to patterning, and patterning would in turn feed back on the expansion dynamics. Our concurrent experimental and modeling work explores this direction in {\it B.\ subtilis} biofilms \cite{us}.

A second natural direction for future work is to extend beyond radial symmetry. Biofilms break radial symmetry in at least two important ways: (1) their edges roughen, acquiring protrusions at characteristic wavelengths \cite{Matsushita1990-uc,Tronnolone2018-kc,Siri2024-tm,Giometto2018-hn}; and (2) on substrates, they grow differently in the directions parallel and perpendicular to the substrate, due to differences in both the mechanical properties and nutrient access along these directions \cite{Serra2015,Yan2017-pk,Bottura2022-tt,Marsden2017-ak,Srinivasan2018-il,Fei2020,Pokhrel2024-dp, Warren2019-um}. For example, recent work has investigated how roughening affects the expansion rate under pressure limitation \cite{Ye2024-sg}. It would be interesting to similarly expand our modeling to investigate both of the these important generalizations.

Our active fluid model is simple and generic. Therefore, we anticipate that the resulting insights will be applicable to active biological systems beyond bacterial biofilms. Examples include tissue dynamics or embryonic development; in both cases, growth is clearly subject to mechanical pressures, nutrient supply, and cell state changes. Our work provides a key template for investigating expansion mechanisms in these and other active matter systems.

\section*{Author Contributions}

All authors designed the research. M.Y.\ performed the research. M.Y.\ and A.M.\ contributed analytic tools. M.Y.\ and A.M.\ analyzed the data. All authors wrote the manuscript.

\section*{Acknowledgments}

M.Y., J.W.L., and A.M.\ were supported by a grant from the NSF-Simons Center for Quantitative Biology. M.Y. and A.M.\ were supported by National Science Foundation grant PHY-2118561 and National Institutes of Health grant R35GM156451. J.M.J. and J.W.L. were supported by a Burroughs Wellcome Fund CASI award and National Institutes of Health grant R35GM142584.

\section*{Appendix A: Numerical solution details}

We solve our model equations numerically by discretizing time and space using the standard Euler method. See \cite{code} for the code.

\begin{figure}[hbt!]
\centering
\includegraphics[width=\linewidth]{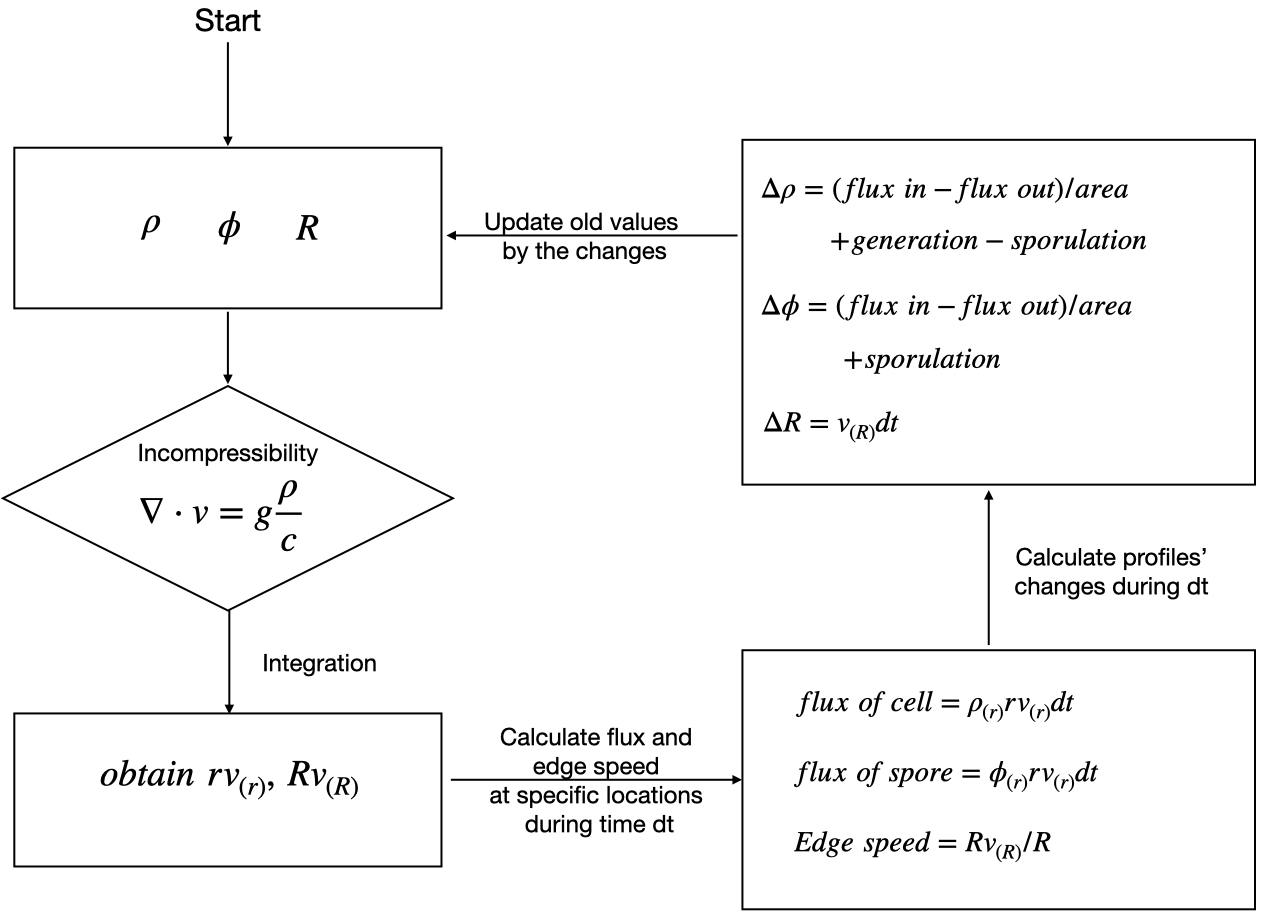}
\caption{Temporal growth arrest numerical solution steps.}
\label{fig:tssimulation}
\end{figure}

For temporal sporulation, because $\rho_1(r,t)$ and $\rho_2(r,t)$ are nonzero within the biofilm radius $R$, the velocity field $v(r,t)$ is valid up to $R$. Therefore, we determine the updated region $R$ by calculating the expansion velocity $v(R,t)$. After determining the updated region, we update the density profile values based on the velocity field and previous profile values. The specific steps are shown in Fig.\ \ref{fig:tssimulation}.

\begin{figure}[hbt!]
\centering
\includegraphics[width=\linewidth]{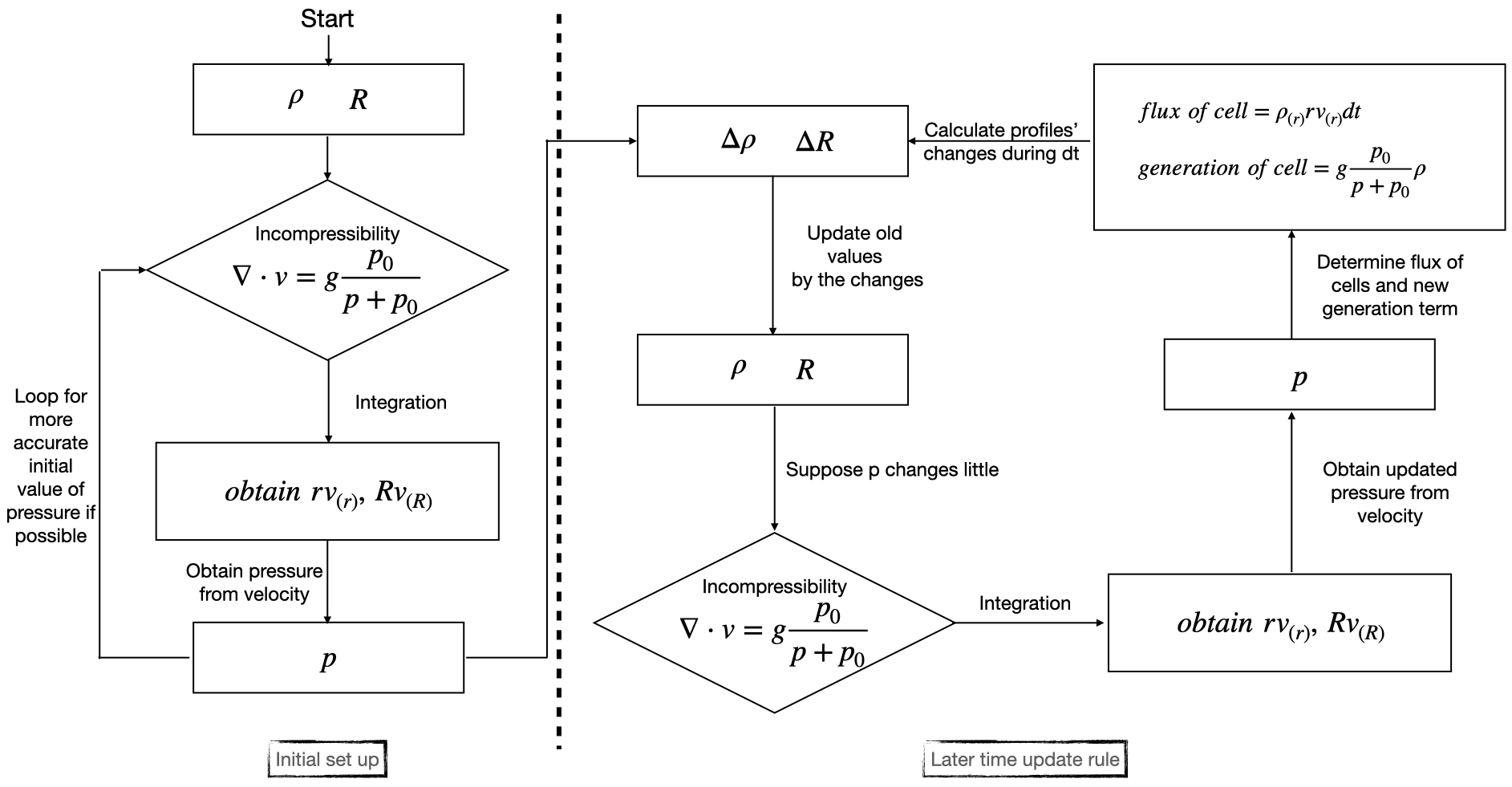}
\caption{Pressure-limited growth numerical solution steps.}
\label{fig:pressuresimulation}
\end{figure}

For pressure-limited growth, as usual the biofilm density is only nonzero within $R$. However, the pressure has a fixed value of $0$ at $R$, and is closely related to velocity. Specifically, velocity is determined by pressure, and pressure is determined by velocity. For simulation efficiency, we update pressure and velocity separately. More specifically, we update velocity and $R$ based on the previous pressure profile. With the new velocity and $R$, we update the pressure. We then use this pressure profile to determine the velocity field in the next iteration. The specific steps are shown in Fig.\ \ref{fig:pressuresimulation}.

\begin{figure}[hbt!]
\centering
\includegraphics[width=\linewidth]{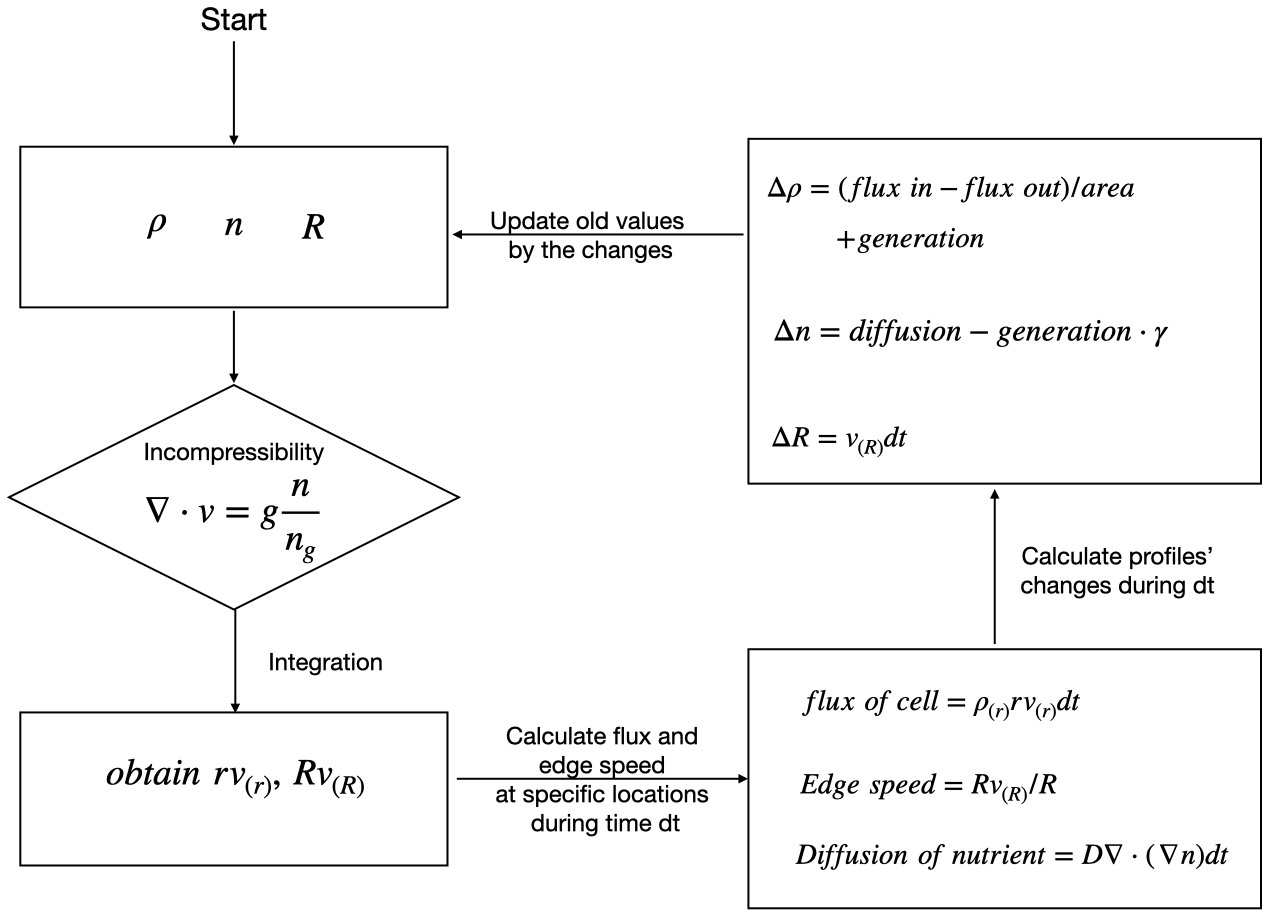}
\caption{Nutrient-dependent growth numerical solution steps.}
\label{fig:nutrientsimulation}
\end{figure}

The steps for nutrient-limited growth are similar to the temporal growth arrest case. Cell density is nonzero within $R$, while nutrients can diffuse in space and can only be converted to cells where cells exist. For efficiency, we implement two other criteria besides total runtime for stopping the solution. The first is that the run ends if the biofilm edge $R$ is sufficiently close to the system radius $L$. The second is that the run ends if $n(r=L)<n_0 R^2/L^2$, indicating the total nutrient left is less than the nutrients initially available for the biofilm and the expansion is slowing down. The specific steps are shown in Fig.\ \ref{fig:nutrientsimulation}.

\section*{Appendix B: Analytic results for 1D and 3D}
Here we extend our analysis to one and to three spatial dimensions. We will find that results are qualitatively the same as in two dimensions.

\subsection*{Temporal growth arrest}
With the active terms in Eq.\ \ref{eq:f1} for temporal sporulation, Eq.\ \ref{eq:dv} with general spatial dimension $d$ reads
\begin{equation}
\label{eq:v_eq1d}
    \frac{\partial}{\partial r}(vr^{d-1})=\frac{g\rho_1r^{d-1}}{c}.
\end{equation}
Integrating over $r$ and solving for $R(t)$ as in Eqs.\ \ref{eq:Rv}-\ref{eq:Rint} gives
\begin{equation}
\label{eq:Rintd}
    R(t) = \left[R_0^d+\frac{gd}{\Omega c}\int_0^tC_1(t')dt'\right]^{1/d},
\end{equation}
where
\begin{equation}
\label{eq:C1d}
    C_1 = \Omega\int_0^{R}\rho_1r^{d-1}dr
\end{equation}
is the total amount of growing cells, and $\Omega = 2\pi^{d/2}/\Gamma(d/2)$ is the angular component of the volume integral (i.e., $\Omega=2, 2\pi, 4\pi, \dots$ for $d=1,2,3,\dots$). Differentiating Eq.\ \ref{eq:C1d} and solving for $C_1(t)$ as in Eqs.\ \ref{eq:dC1}-\ref{eq:C1t} gives
\begin{equation}
\label{eq:C1td}
    C_1(t) = C_1(0)e^{(g-k)t},
\end{equation}
making Eq.\ \ref{eq:Rintd}
\begin{equation}
\label{eq:R1d}
    R(t) = \left\{R_0^d+\frac{d}{\Omega}\frac{C_1(0)}{c}\frac{g}{g-k}\left[e^{(g-k)t}-1\right]\right\}^{1/d}.
\end{equation}
At sufficiently long times we have $R(t) \sim e^{(g-k)t/d}$, or explicitly in one and three dimensions,
\begin{equation}
    R(t) \sim
    \begin{cases}
    e^{(g-k)t} & {\rm 1D}, \\
    e^{(g-k)t/3} & {\rm 3D}.
    \end{cases}
\end{equation}
We see that, as in two dimensions, growth must outpace sporulation ($g>k$), and sporulation slows the expansion rate but does not prevent the biofilm from spreading exponentially.

\begin{figure*}[hbt!]
\centering
\includegraphics[width=0.8\linewidth]{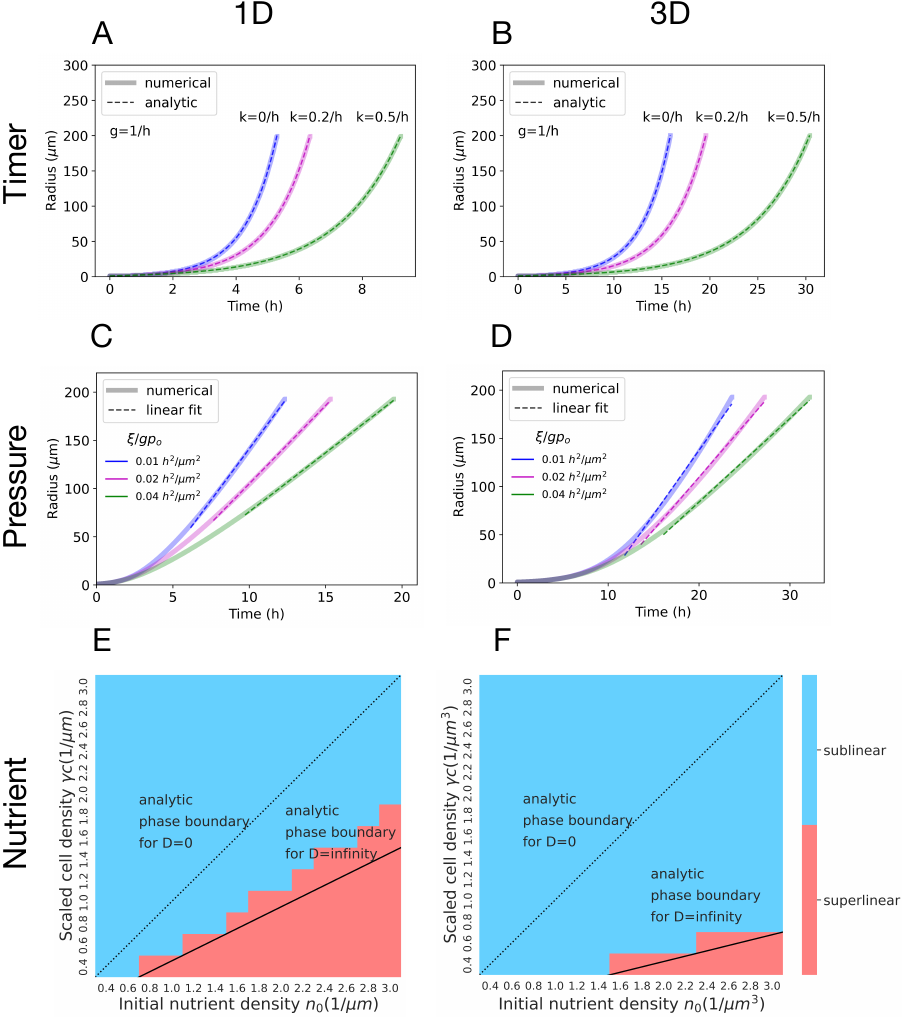}
\caption{Spreading dynamics in 1D (left) and 3D (right) are qualitatively similar for the three cases to those in 2D in the main text. Parameters used are as in Fig.\ \ref{fig:2dtimer simulation result} (A and B), Fig.\ \ref{fig:2dpressureresult} (C and D), and Fig.\ \ref{fig:2dnutrientsimulation_small_l} (E and F).}
\label{fig:1d3dsimulationresult}
\end{figure*}

The numerical results and analytics are shown in Fig.\ \ref{fig:1d3dsimulationresult}A for one dimension and Fig.\ \ref{fig:1d3dsimulationresult}B for three dimensions.

\subsection*{Pressure-limited growth}
With the active term in Eq.\ \ref{eq:f2} for pressure-limited growth, Eq.\ \ref{eq:dv} with general spatial dimension $d$ reads
\begin{equation}
\label{eq:v_eq2app}
    \frac{1}{r^{d-1}}\frac{\partial}{\partial r}(r^{d-1}v)=\frac{gp_0}{p+p_0}.
\end{equation}
Differentiating with respect to $r$ and assuming $p/p_0\ll 1$ as in Eqs.\ \ref{eq:v2order_edited}-\ref{eq:bessel} gives
\begin{equation}
\label{eq:besseld}
    x^2\frac{\partial^2 v}{\partial x^2} + (d-1)x\frac{\partial v}{\partial x} -(x^2+d-1)v = 0,
\end{equation}
with $x=r/\lambda$ and $\lambda = \sqrt{p_0/(g\xi)}$. Equation \ref{eq:besseld} satisfies the boundary conditions
\begin{align}
\label{bc1}
v(0,t) &= 0, \\
\label{bc2}
\frac{(d-1)v(R,t)}{R}+\left.\frac{\partial v}{\partial r}\right|_R &= g,
\end{align}
where the second condition comes from evaluating Eq.\ \ref{eq:v_eq2app} at $r=R$ where $p=0$, as in Eq.\ \ref{pbc}.

For $d=1$, Eq.\ \ref{eq:besseld} reduces to
\begin{equation}
\label{eq:bessel_1d}
    \frac{\partial^2 v}{\partial x^2} -v = 0,
\end{equation}
which is solved by
\begin{equation}
\label{eq:vAB_1d}
    v(x,t) = A(t)e^x + B(t)e^{-x}.
\end{equation}
The boundary conditions (Eqs.\ \ref{bc1} and \ref{bc2}) determine $A$ and $B$. Evaluating the resulting expression for $v$ at $r=R$ gives
\begin{equation}
\label{eq:Rode2_1d}
    v(R,t)=\frac{dR}{dt}=\frac{g\lambda(e^{R/\lambda}-e^{-R/\lambda})}{e^{R/\lambda}+e^{-R/\lambda}}.
\end{equation}
At long times, when $R\gg\lambda$,
\begin{equation}
\label{eq:linear_speed_1d}
    \frac{dR}{dt} \to g\lambda= \sqrt{\frac{p_0g}{\xi}}
    \qquad {\rm 1D}.
\end{equation}

For $d=3$, Eq.\ \ref{eq:besseld} is solved by
\begin{equation}
\label{eq:vAB_3d}
    v(x,t) = A(t)i_1(x) + B(t)k_1(x)
\end{equation}
where $i_1(x)$ and $k_1(x)$ are first-order modified spherical Bessel functions. Again the boundary conditions (Eqs.\ \ref{bc1} and \ref{bc2}) determine $A$ and $B$, as in Eqs.\ \ref{eq:vAB}-\ref{eq:Rode2}, and evaluating the resulting expression for $v$ at $r=R$ gives
\begin{equation}
\label{eq:Rode23}
    v(R,t)=\frac{dR}{dt} = \frac{g\lambda Ri_1(R/\lambda)}{2\lambda i_1(R/\lambda) + Ri'_1(R/\lambda)},
\end{equation}
where $i_1'(x) = di_1/dx$. For large argument, $i_1(x) \to e^x/(2x)$ and therefore $i'_1(x) \to i_1(x)[1-1/x] \approx i_1(x)$. Thus, for $R\gg \lambda$, the Bessel functions in Eq.\ \ref{eq:Rode23} cancel, and the $2\lambda$ in the denominator can be neglected, giving
\begin{equation}
\label{eq:linear_speed_3d}
    \frac{dR}{dt} \to g\lambda= \sqrt{\frac{p_0g}{\xi}}
    \qquad {\rm 3D}.
\end{equation}
Equations \ref{eq:linear_speed_1d} and \ref{eq:linear_speed_3d} are identical to the two-dimensional result, Eq.\ \ref{eq:linear_speed}. In all cases, the expansion speed approaches a constant at long times.

The numerical results and analytics are shown in Fig.\ \ref{fig:1d3dsimulationresult}C for one dimension and Fig.\ \ref{fig:1d3dsimulationresult}D for three dimensions.

\subsection*{Nutrient-limited growth}
With the active term in Eq.\ \ref{eq:f3} for nutrient-limited growth, Eq.\ \ref{eq:dv} with general spatial dimension $d$ reads
\begin{equation}
\label{eq:v_eq3d}
    \frac{\partial}{\partial r}(vr^{d-1})=\frac{gnr^{d-1}}{n_g}.
\end{equation}
Integrating over $r$ and solving for $R(t)$ as in Eqs.\ \ref{eq:Rv3}-\ref{eq:Rint3} gives
\begin{equation}
\label{eq:Rint3d}
    R(t) = \left[R_0^d+\frac{gd}{\Omega n_g}\int_0^tN(t')dt'\right]^{1/d},
\end{equation}
where
\begin{equation}
\label{eq:Nd}
    N = \Omega\int_0^{R}nr^{d-1}dr
\end{equation}
is the total amount of nutrient within the biofilm, and again $\Omega = 2\pi^{d/2}/\Gamma(d/2)$.
Differentiating Eq.\ \ref{eq:Nd} as in Eqs.\ \ref{eq:dN1}-\ref{eq:dN3_edited} gives
\begin{equation}
\label{eq:dN3_editedd}
    \frac{dN}{dt}
    = \frac{g}{n_g}[n(R,t)-\gamma c]N + \Omega DR^{d-1}\left.\frac{\partial n}{\partial r}\right|_R.
\end{equation}
For either $D=0$ or $D\to\infty$, the second term in Eq.\ \ref{eq:dN3_editedd} vanishes: for $D=0$ it vanishes directly, and for $D\to\infty$ we assume that $(dn/dr)|_R$ vanishes more strongly than $D$ diverges, as in the main text. In both cases, Eq.\ \ref{eq:dN3_editedd} becomes
\begin{equation}
\label{eq:dN3_editedd2}
    \frac{dN}{dt}
    = \frac{g}{n_g}[n(R,t)-\gamma c]N.
\end{equation}

For $D=0$, we have $n(R,t) = n_0$, and the qualitative dynamics of $N$, and thus $R$, will depend on the sign of $n_0-\gamma c$, as in Eqs.\ \ref{eq:dN4}-\ref{eq:phase}. Therefore, the phase boundary for $D=0$ does not depend on spatial dimension $d$. Explicitly, in one and three dimensions, the phase boundary is given by
\begin{equation}
D=0:
\begin{cases}
    n_0 = \gamma c & {\rm 1D},\\
    n_0 = \gamma c & {\rm 3D}.
\end{cases}
\end{equation}

For $D\to\infty$, we find the acceleration $a=d^2R/dt^2$ as in Eqs. \ref{d2R} and \ref{a1}, giving
\begin{equation}
\label{ad}
a = \frac{g^2N}{n_g^2\Omega R^{d-1}}\left[n-\gamma c-\frac{(d-1)N}{\Omega R^d}\right].
\end{equation}
The generalization of Eqs.\ \ref{eq:nu_inf_n2_new} and \ref{NRD} for the uniform nutrient concentration is
\begin{equation}
\label{nd}
n = \frac{N}{\Omega R^d/d} = \frac{n_0L^d-\gamma c(R^d-R_0^d)}{L^d},
\end{equation}
where $\Omega/d$ is the prefactor for the volume of a $d$-dimensional sphere.
Inserting Eq.\ \ref{nd} into Eq.\ \ref{ad} gives
\begin{equation}
    a = \frac{g^2N}{n_g^2\Omega R^{d-1}}\left[\frac{n_0}{d}-\frac{\gamma c}{d}\left(\frac{R^d-R_0^d}{L^d}\right)-\gamma c\right],
\end{equation}
or, for $R\to L\gg R_0$,
\begin{equation}
    a = \frac{g^2N}{n_g^2\Omega R^{d-1}d}\left[n_0-(d+1)\gamma c\right].
\end{equation}
We see that $a$ changes sign at the phase boundary $n_0 = (d+1)\gamma c$. Explicitly, in one and three dimensions, the phase boundary is given by
\begin{equation}
D\to\infty:
\begin{cases}
    n_0 = 2\gamma c & {\rm 1D},\\
    n_0 = 4\gamma c & {\rm 3D}.
\end{cases}
\end{equation}

The numerical results and analytic phase boundaries are shown in Fig.\ \ref{fig:1d3dsimulationresult}E for one dimension and Fig.\ \ref{fig:1d3dsimulationresult}F for three dimensions. There we use the small system size parameters of Fig.\ \ref{fig:2dnutrientsimulation_small_l}, where the $D\to\infty$ limit is expected to hold.

\end{document}